\documentclass[aip, jmp, amsmath, amssymb, reprint]{revtex4-1}
\usepackage{graphicx}
\usepackage{fullpage}
\usepackage{color}
\usepackage[normalem]{ulem} %REMOVE AFTER REVISIONS
\begin{document}
\title{Phase effects due to previous pulses in time-resolved Faraday rotation measurements}
\author{Christopher J. Trowbridge}\affiliation{Department of Physics, University of Michigan, Ann Arbor, MI 48109}
\author{Vanessa Sih}\affiliation{Department of Physics, University of Michigan, Ann Arbor, MI 48109}
\date{\today}
\begin{abstract}
Time-resolved Faraday rotation measurements have proved transformative in the investigation of spin dynamics in semiconductors.  In materials with spin lifetimes which are on the order of, or greater than, the laser repetition time, the collective effect of spin polarization due to the whole pump pulse train becomes important.  Here, we discuss a relative phase shift which results from these spins. We derive and experimentally validate a closed-form expression which describes this phase shift, and characterize it throughout parameter space. A spin lifetime measurement based on this phase shift is described, and we discuss situations in which the model used must be augmented to be applicable.
\end{abstract}
\maketitle

\section{Introduction}
Time-resolved Faraday rotation (TRFR) measurements have proved to be a powerful tool for the study of dynamical spin phenomena in semiconductors \cite{baumberg_1994, crooker_1995, crooker_1996, kikkawa_1997}.  In its most basic form, TRFR is an optical pump-probe technique that can be used to monitor the temporal evolution of an optically-injected spin polarization and determine the electron g-factor and spin lifetime in materials with advantageous selection rules\cite{optical_orientation}.  Variations on TRFR measurements, including spin-drag and Larmor magnetometry, enable the study of spin transport properties \cite{stepanov_2014,norman_2014}, spin-orbit fields\cite{norman_2010,kuhlen_2012}, dynamic nuclear polarization\cite{kikkawa_2000,kawakami_2001,trowbridge_2014}, and many other phenomena.  When the spin polarization lifetime $T_2^*$ is on the order of or larger than the laser repetition time, the spins excited by previous pump pulses make non-negligible contributions to the Faraday rotation signal, resulting in resonant spin amplification (RSA)\cite{kikkawa_1998,glazov_2008,yugova_2012}.  In samples with long lifetimes, RSA can provide a more accurate way to measure $T_2^*$ than TRFR\cite{kikkawa_1998}.  In this paper, we investigate another effect of previous pump pulses which results in an effective phase shift that depends on the applied magnetic field.  This phase shift must be accounted for in any situation in which the contribution to TRFR from previous pulses is expected to be important, and can provide another method for determining the spin lifetime.

\section{TRFR Measurements}
In a typical TRFR measurement, a circularly polarized laser pulse excites a spin polarization parallel to the direction of laser propagation, which we take to be $\hat{z}$. After some time delay, a linearly polarized probe pulse interrogates the system by measuring Faraday rotation, which is proportional to the spin polarization along $\hat{z}$.  Typically, an ultrafast laser is used to generate the pulses and the time delay between pulses is controlled with a mechanical delay line. In this way, the shortest accessible time scale is set by the pulse time and the precision with which the delay line can be positioned. The spin polarization lifetime can be found by measuring the timescale of the decay of Faraday rotation as $\Delta t$, the delay between pump and probe pulses, is swept.  For a measurement of the g-factor, a magnetic field is applied in the Voigt geometry along $\hat{x}$. Spins will then precess in the $\hat{y}$-$\hat{z}$ plane at their Larmor frequency $\Omega_L = g \mu_B B/\hbar$,  where g is the electron g factor, $\mu_B$ is the Bohr magneton, and $\hbar$ the reduced Planck constant. In this geometry, a single pump pulse gives rise to a spin polarization of: 
\begin{equation}
s_{\hat{z}} = A\cos(\Omega_L \Delta t)\exp[(-\Delta t)/T_2^*]\Theta(\Delta t)
\end{equation}
where A is the initial spin polarization and $\Theta(\Delta t)$ is the Heaviside step function which enforces causality in the expression.  

When the spin lifetime is such that an appreciable spin polarization remains after the laser repetition time, spins due to previous pump pulses must be taken into account as well.  It is convenient at this point to use phasor notation, $\vec{s}\rightarrow\underbar{s}$ where $\underbar{s}$ is a complex spin phasor. In this notation, $s_{\hat{z}}$ is understood to be the real part of $\underbar{s}$, Re$(\underbar{s})$ and $s_{\hat{y}}$ becomes the imaginary part of $\underbar{s}$,  Im$(\underbar{s})$. In this case, each pump pulse contributes:
\begin{equation}
\underbar{s} = A\exp\left[(i\Omega_L - 1/T_2^*)\Delta t\right]\Theta(\Delta t)
\end{equation}
 to the total spin. Summing over previous pulses, the total $\underbar{s}$ becomes: 
\begin{eqnarray}
\underbar{s} = \sum_{n} A&\exp& \left[(i \Omega_L -1/T_2^*)(\Delta t + nT_{rep}) \right]\nonumber\\
	&\times&\Theta(\Delta t + nT_{rep})
\end{eqnarray}
where $T_{rep}$ is the laser repetition time.

\subsection{Experimental data}

Figure 1 (a) shows experimental data taken on a commercially purchased sample consisting of a 2 $\mu$m layer of n-GaAs doped at $1\times 10^{17}$ cm${}^{-3}$ atop a semi-insulating GaAs substrate.  Measurements were performed at 10 K as a function of external magnetic field $B_{ext}$ and pump-probe delay time $\Delta t$. A mode-locked Ti:Sapphire laser tuned to $\lambda = 818.2$ nm with a pulse time of approximately 3 ps and a repetition time $T_{rep}$ of 13.2 ns was used.  Pump and probe beam powers were measured to be 500 $\mu$W and 75 $\mu$W, respectively, with beam diameters of approximately 35 $\mu$m each.  Here, the magnitude of the signal shown is in millivolts as measured on a lock-in amplifier, which is proportional to $\theta_F$, the Faraday rotation angle, but taken to be an arbitrary unit.  The data set was then fit simultaneously to extract the electron g factor and spin lifetime.  The data are fit to the real part of Eq. (3), with a term added which takes into account the contribution from optically excited carriers before recombination, which decays rapidly.  In total, the fit equation used is: 
\begin{equation}
\begin{split}
\theta_f=Re\Big\{\sum_{n}&A\exp[(i\Omega_L-1/T_2^*)(\Delta t + nT_{rep})]\\
&\times\Theta(\Delta t + nT_{rep})\Big\}\\
+ A_O\cos&(\Omega_L\Delta t)\exp[\Delta t/T_O]\times\Theta(\Delta t)
\end{split}
\end{equation}
The fit finds $g =  -0.3788$\cite{footnote}, $T_2^* = 10.2$ ns, and optical spin lifetime $T_O = 87$ ps.  The ratio of Faraday rotation from long-lived spins to spins from optical carriers, $A/A_O$, is found to be $2.0$.  Figure 1 (b) shows computed data using the results of the fit. Both plates in Fig. 1 have been normalized using the same normalization factor. 

\begin{figure} 
\includegraphics[width = 3.125 in]{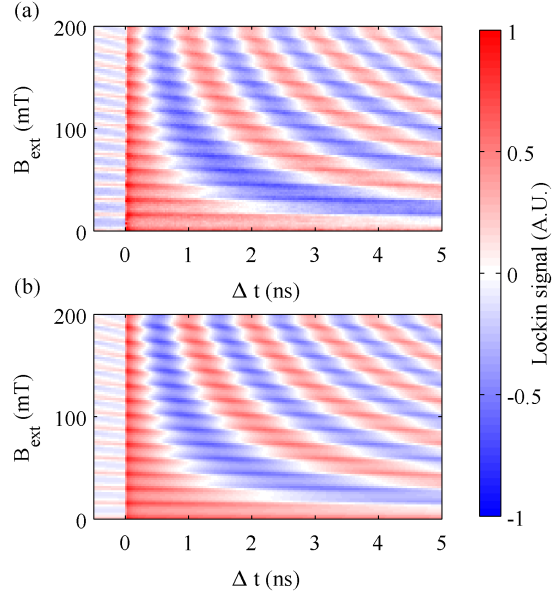}
\caption{Normalized lock-in signal (proportional to Faraday rotation angle $\theta_f)$ as a function of delay time $\Delta t$ (horizontal axis) and external magnetic field $B_{ext}$ (vertical axis) for experimental data (a) and data generated using Eq. (4) with fit parameters from experimental data (b).}
\end{figure}

\subsection{Resonant Spin Amplification}
Measurements which rely on resonant spin amplification take advantage of the collective effect of spin polarization resulting from previous pulses.  Typically, the external magnetic field is scanned instead of the delay time.  When the magnetic field reaches a value at which spins precess an integer number of times in the laser repetition time, spins resulting from all previous pump pulses are aligned along $\hat{z}$ when the next pulse arrives, resulting in a large spin polarization at $\Delta t = 0^+$. When a half-integer number of precessions occur, spins antialign and a smaller polarization is seen. This is shown diagramatically in Fig. 2 plates (a) and (b).  These figures show the $\hat{y}$ and $\hat{z}$ components of spin polarization at $\Delta t = 0^+$, immediately after a pump pulse arrives, normalized to the initial spin polarization of a single pulse $A$.  Contributions to the total polarization from five additional pump pulses are shown in blue, along with the total spin polarization in red. Here, data was synthetically generated using $g = -0.3788$, $T_{rep} = 13158$ ps and $T_2^* = 2 T_{rep}$.  

\begin{figure}
\includegraphics[width = 3.125 in]{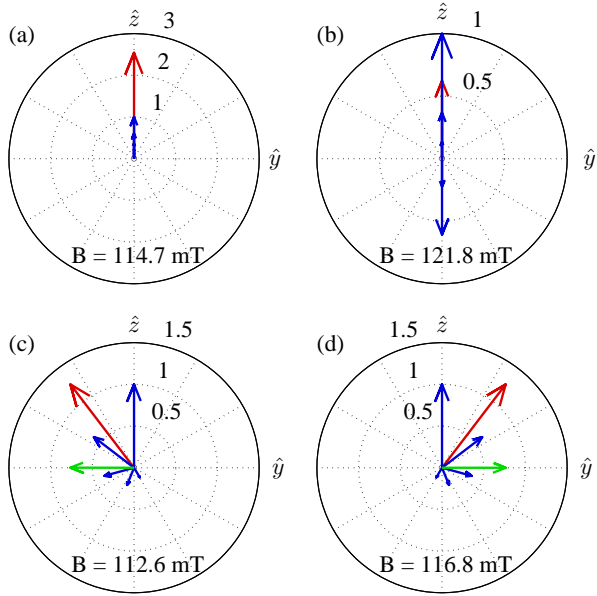}
\caption{Orientation of total spin polarization (red) and contributions due to the five most recent pump pulses (blue) for maximum and minimum amplitude [(a) and (b), respectively] and minimum and maximum phase [(c) and (d), respectively] immediately following the arrival of a pump pulse ($\Delta t = 0^+$).  In plates (c) and (d), the green arrow indicates the sum of all previous pulses, excluding the most recent pulse. Data generated using Eq. (3), with $T_2^* = 2T_{rep}$.  All vectors are normalized to $A = 1$, and inset numbers label lines of constant radius.}
\end{figure}

\begin{figure}
\includegraphics[width = 3.125 in]{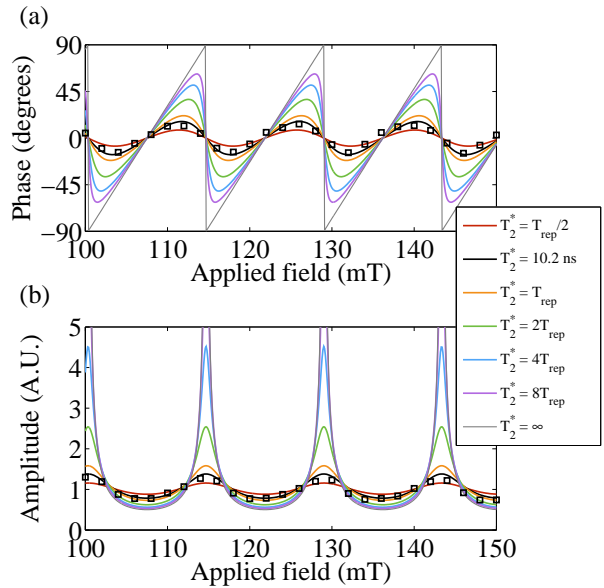}
\caption{Effective phase $\phi$ (a) and amplitude $r$ (b) as defined in Equations (7) and (8) from model data with a range of lifetimes (solid lines) and experimental data (black squares).}
\end{figure}

The degree to which the spin polarization at $\Delta t=0^+$ exceeds A when a resonance condition is met is determined by the lifetime of the spins; longer lifetimes result in higher and sharper resonance peaks as measured by scanning the external magnetic field.  The g-factor and lifetime can be found by fitting this data to Eq. 3.  This method can give better fits of the lifetime than standard TRFR scans when it is large compared to the accessible delay times\cite{kikkawa_1998}.

\section{Phase shift due to previous pulses}
By pulling terms out of the sum in Eq. (3) which do not depend on $n$ and enforcing $\Delta t \geq 0$,  $\underbar{s}$  becomes: 
\begin{eqnarray}
\underbar{s} = A\exp[(&i&\Omega_L  - 1/T_2^*)\Delta t]\nonumber\\
\times&&\sum_{n\geq 0}\exp[(i\Omega_L  - 1/T_2^*)nT_{rep}]
\end{eqnarray}
where now the sum is in the form of a converging geometric series with common ratio $\exp[(i\Omega_L-1/T_2^*)T_{rep}]$.  The sum can therefore be evaluated: 
\begin{equation}
\sum_{n\geq 0}\left(e^{i\Theta-x}\right)^n = \frac{1}{1-e^{i\Theta-x}} 
\end{equation}
with $\Theta = \Omega_LT_{rep}$ and $x = T_{rep}/T_2^*$.  This can be rewritten as a complex number $re^{i\phi}$ whose amplitude $r$ and phase $\phi$ are functions of the external magnetic field and spin lifetime:
\begin{eqnarray}
r = \left(1-2e^{-x}\cos(\Theta) + e^{-2x}\right)^{-\frac{1}{2}}\\
\phi = \tan^{-1}\left(\frac{e^{-x}\sin(\Theta)}{1-e^{-x}\cos(\Theta)}\right)
\end{eqnarray}
By taking the real part of Eq. (5) using these expressions, $s_{\hat{z}}$ becomes: 
\begin{equation}
s_{\hat{z}} = rA\cos(\Omega_L\Delta t +\phi)e^{-\Delta t/T_2^*}
\end{equation}
These expressions are functionally equivalent to the closed-form expression found in Ref. 17, but are valid for $\Delta t\in[0, T_{rep})$ instead of $\Delta t\in[-T_{rep},0)$, and are written in a form that makes explicit the fact that previous pulses contribute an effective phase shift $\phi$ to TRFR measurements in addition to modifying the amplitude of the spin polarization. This phase shift is equivalent to $\tan^{-1}(s_{\hat{y}}/s_{\hat{z}})\rvert_{\Delta t = 0^+}$.

In an effect analogous to peaks in RSA amplitude, there exist values of the external field where the sum of the spin polarization due to previous pulses will result in a maximum effective phase.  At these field values, the spin polarization immediately preceding the arrival of a pump pulse $\vec{s}(0^-)$, will lay in the sample plane along $\hat{y}$\cite{supplement}.  This condition is shown in Fig. 2 plates (c) and (d).  In these figures, the green vector indicates the sum of previous pulses excluding the most recent pulse. Again, the red and blue arrows show the total spin polarization and the first five terms in the sum, respectively.  At field values which satisfy this condition, the effective phase $\phi$ in Eq. (9) is maximized.  

\begin{figure}
\includegraphics[width = 3.125 in]{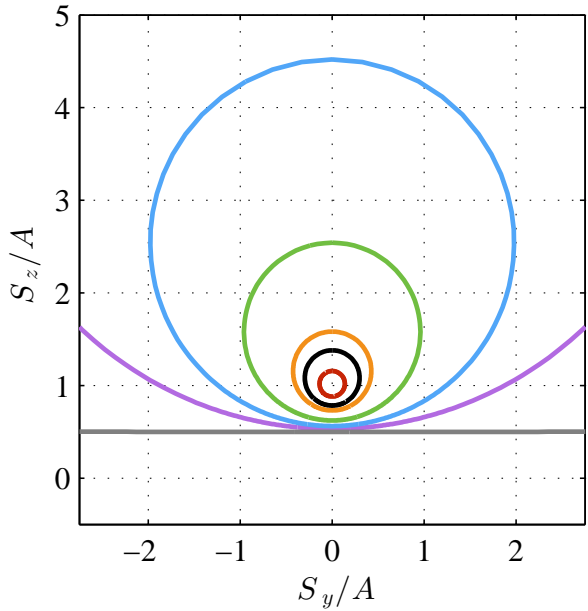}
\caption{Path traced out by total spin polarization $\vec{s}(\Delta t = 0^+)$ as applied magnetic field is swept.  In the limit $T_2^*\rightarrow\infty$, the radius of the circle generated goes to infinity, which results in maximum and minimum phases of $\pi/2$ and $-\pi/2$, respectively. }
\end{figure}

Figure 3 shows plots of $r$ and $\phi$ versus external magnetic field calculated using Equations  (7) and (8) with a range of lifetimes.   Individual delay scans taken from the experimental data in Fig. 1 (a) were fit to Eq. (9) to extract $r$ and $\phi$, and the results are plotted as black squares. Calculated values of $r$ and $\phi$ using the experimentally determined lifetime and g factor from the simultaneous fit above are shown as a black line to highlight the close agreement with the model.  As the spin lifetime increases, the maximum and minimum observed phase angles increase in magnitude and are found closer to the external magnetic field values corresponding to an RSA peak.  

As the magnetic field is swept, the total spin polarization at $\Delta t = 0^+$, depicted by the red arrows in Fig. 2, traces out a circular path in the $\hat{y}$-$\hat{z}$ plane.  This path is shown in Fig. 4 for the lifetimes used previously in Fig. 3.  As the lifetime grows, the radius of the circle increases, and in the limit that $T_2^*$ goes to infinity the radius of this circle diverges.  Accordingly, the maximum and minimum phase angles approach $\pi/2$ and $-\pi/2$, respectively. The rate at which $\vec{s}(\Delta t = 0^+)$ moves around the circle as the magnetic field is swept is not constant, but rather exhibits a rapid shift in phase as the field value moves past an RSA peak and then a slow shift in phase through the RSA minimum.  

\subsection{Measurements using $\phi$}
Since arctangent grows monotonically with its argument, the extrema of $\phi$ can be found by extremizing the argument of the arctangent in Eq. (8). Setting its derivative equal to zero and solving gives the values of $\Theta$ which extremize $\phi$: 
\begin{equation}
\Theta_{extr} = 2n\pi \pm\cos^{-1}(e^{-x})
\end{equation}
As $T_2^*$ grows (causing $x$ to shrink with constant $T_{rep}$), the values of $\Theta_{extr}$ grow closer to $2n\pi$, which correspond to peaks in $r$.  Substituting Eq. (10) with $n$ = 0 into Eq. (8) gives the maximum phase angle as a function of $x$, given by:
\begin{equation}
\phi_{max} = \tan^{-1}\left(\frac{e^{-x}}{\sqrt{1-e^{-2x}}}\right)
\end{equation}
which is plotted in Fig. 5 (a).  $\phi_{max}$ has a monotonic dependence on the spin lifetime, and can therefore be used to measure $T_2^*$ when it is sufficiently large compared to the laser repetition time. Such a measurement could be performed by first setting $\Delta t$ to a small negative value and tuning the magnetic field so that the Faraday rotation disappears.  At this magnetic field, the polarization due to previous pulses will lay in the sample plane, and correspond to a condition of maximum or minimum phase.  Then, a delay scan would be performed and fit to extract $\phi$, which, by inverting Eq. (11), can be used to find $T_2^*$.  In the experimental data shown in Fig. 3, the maximum observed phase angle was $16.4^\circ$, which gives a lifetime of 10.4 ns, close to the value obtained above.  

\begin{figure}
\includegraphics[width = 3.125 in]{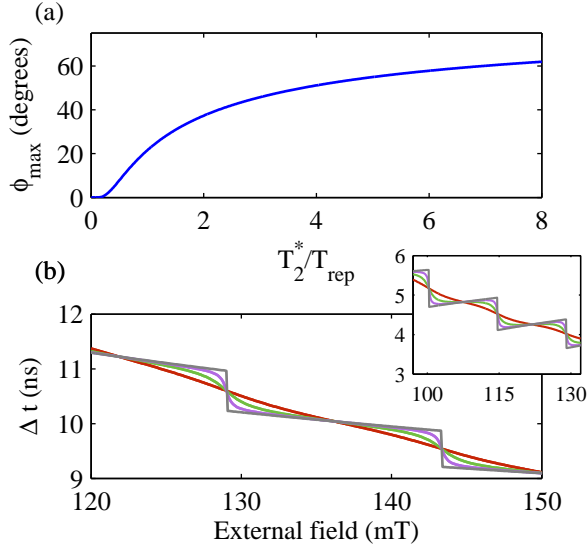}
\caption{(a) Maximum phase angle $\phi_{max}$ versus $T_2^*/T_{rep}$, calculated using Eq. (10).  (b) Delay time corresponding to the $15^{th}$ zero crossing of $s_{\hat{z}}$ (vertical axis) for a subset of the lifetimes defined in the Fig. 3 legend versus external magnetic field.  (Inset) Zero crossing delay time for the $6^{th}$ crossing.  In this range of external magnetic field strength, a breakdown of the one-to-one correspondence between $\Delta t_0$ and $B$ occurs for long spin lifetimes.}
\end{figure}

TRFR can be used to measure the local magnetic field when the g factor is known by a process dubbed Larmor magnetometry\cite{kikkawa_2000}. Traditionally, a delay scan is performed over at least a few periods of precession and a fit to the precession frequency extracts the magnetic field present.  The speed with which such a measurement can be performed is set by the time it takes to complete the delay scan, typically on the order of 10 to 100 seconds.  Measurements have also been performed by maintaining a fixed delay and continuously monitoring $\theta_F$ \cite{kawakami_2001}.  This technique does not generally allow for a good determination of the magnetic field because $\theta_F$ is not single-valued when the magnetic field changes, and when near a local extrema, $\theta_F$ changes slowly with magnetic field, reducing or eliminating the ability to measure changes in the magnetic field. Additionally, shifts in laser power or slow drifts in pump-probe overlap are indistinguishable from changes in magnetic field.  These limitations can be avoided if instead the delay time is continuously adjusted to follow a zero crossing of $\theta_F$ of known order. The position at which a zero crossing occurs is unaffected by changes in laser power or pump-probe overlap, and, with care, a zero crossing can be picked whose delay time has a one-to-one correspondence with the magnetic field strength over a range of values.  A measurement can then be completed in the time it takes to adjust the position of the mechanical delay line once and take a measurement of $\theta_F$, which can take place in approximately one second.

A zero crossing will occur when $s_{\hat{z}} = 0$, or from Eq. (9), the $n^{th}$ crossing occurs when $\Omega_L\Delta t + \phi = (2n-1)\pi/2$.  The presence of $\phi$ in this expression complicates this measurement.  Figure 5 (b) shows the delay time at which the $n=15$ crossing occurs versus magnetic field for a range of lifetimes.  Again, colors refer to the legend in Fig. 3.  When the spin lifetime is long, the change in phase due to previous pulses is large, and results in a series of sharp steps in the delay time at which a zero crossing occurs that correspond with RSA peaks as the magnetic field strength is scanned.  It can be shown that if a zero crossing occurs at a delay time that satisfies $\Delta t > T_{rep}/2$, there will always be a one-to-one relationship between the local magnetic field strength and the delay time at which the crossing occurs.  The inset in Fig. 5 (b) shows what happens when this relation is violated for a subset of the spin lifetimes defined in Fig. 3.  Here, the $6^{th}$ crossing has been chosen.  In this situation, for long spin lifetimes, the local magnetic field can no longer be uniquely determined from the delay time of the zero crossing.  In order to measure the local magnetic field using this technique, the spin lifetime must now be known, which adds uncertainty to the measurement of the magnetic field.  Additionally, it must be known exactly which crossing is being followed.  This could be ascertained by performing a full delay scan at a fixed, known magnetic field.

\subsection{Breakdown of the model} 
In the preceding analysis, it is assumed that the contributions from previous pulses decay exponentially with the time elapsed since the pump pulse arrived at the sample.  In situations where there is electron drift or diffusion, each term in the sum in Eq. (3) must now include a factor which takes this into account.  In general, this will depend on many parameters, including but not limited to beam size and shape, mobility, in-plane electric fields, and diffusion constant. Equation (5) becomes: 
\begin{eqnarray}
\underbar{s} = \sum_{n} \big\{A\exp[(&i& \Omega_L -1/T_2^*)(\Delta t + nT_{rep}) ]\nonumber\\
\times &&F_n(\Delta t)\Theta(\Delta t + nT_{rep})\big\}
\end{eqnarray}
where $F_n(\Delta t)$ takes into account changes in overlap.  Since $F_n(\Delta t)$ depends on $n$ and $\Delta t$, it cannot be factored out of the sum.  As a result, this expression cannot be simplified to the form of Eq. (9), and the results of a fit to that functional form will no longer directly give the precession frequency or lifetime. Care must be taken to characterize and correct for this effect whenever the in-plane motion of electrons is expected to be important.  Also, we assume that spins resulting from each pump pulse are not affected by subsequent pulses.  If the optical power is sufficient that the number of optical carriers is on the order of or greater than the number of dopant carriers within the beam area, this is no longer a good assumption\cite{kuhlen_2014}, and a factor which accounts for this must be included in the sum.  This factor will not depend on $\Delta t$, however, so that Eq. (9) may still be valid, though the expressions for $r$ and $\phi$ must be revised.  This effect can be minimized by decreasing the pump and probe powers.

\section{Conclusion}
In conclusion, we have recast expressions used in the context of RSA measurements in a form that highlights an effective phase shift due to the collective action of spin polarization from previous pulses in TRFR measurements. We experimentally measure this phase shift in n-GaAs and find that it is in good agreement with the model.  In situations with long spin lifetimes, the phase shift can become a significant factor in TRFR measurements.  We suggest a measurement which utilizes the phase to provide another means of accessing the spin lifetime when it is long compared to the laser repetition time in which $\Delta t$ is scanned instead of the external magnetic field. 

This work was supported by the National Science Foundation Materials Research Science and Engineering Center program DMR-1120923 and under Grant No. ECCS-0844908, the Office of Naval Research, the Air Force Office of Scientific Research, and the Defense Threat Reduction Agency Basic Research Award No. HDTRA1-13-1-0013.


\begin{thebibliography}{99}
\bibitem{footnote} The sign of the g-factor is not accessible to this measurement, but is assumed to be negative in accordance with values found in the literature.
\bibitem{supplement} This can be shown by evaluating $s_{\hat{z}}(\Delta t = 0^+)$ with $\Theta$ given by Eq. (10) and subtracting off the contribution from the most recent pulse, which gives $s_{\hat{z}}(\Delta t = 0^-) = 0$.
\end{thebibliography}
\end{document}